\documentstyle[epsfig]{elsart}              

\newcommand{\bmf}[1]{\mbox{\boldmath $#1$}}

\begin{document}
\begin{frontmatter}

                                                                              
\title{                                                                         
Determination of the correlation length 
from short-time dynamics} 

\author{A.~Jaster}
\address{Universit\"{a}t - GH Siegen, D-57068 Siegen, Germany}
\date{\today}
                                                                                

\maketitle
\begin{abstract}

We investigate the short-time dynamic relaxation of the two-dimensional 
XY model in the high temperature phase. Starting from the ordered state,
we measure the autocorrelation function and determine the 
autocorrelation time. It is shown that apart from a constant 
the equilibrium 
spatial correlation length can be calculated in this way.
\end{abstract}

\keyword
Short-time dynamics; Non-equilibrium kinetics; Monte Carlo simulation;
spin model
\PACS{82.20.Mj; 02.70.Lq; 75.10.Hk; 75.40.Gb}
\endkeyword

\end{frontmatter}

Simulations in the short-time regime \cite{ZHENG}
gain more and more importance since Janssen, Schaub and Schmittmann
\cite{JASCSC} showed that universality exists already in this early time of the
evolution. Traditionally, it was believed that universal scaling behaviour can
be found only in the long-time regime. However,
Janssen et.\ al discovered that a magnetic system  with non-conserved
order parameter (model A \cite{HOHHAL}) quenched from a high temperature state
to the critical temperature shows universal scaling behaviour already at
short times. After a microscopic time $t_{\mathrm{mic}}$,
the magnetization undergoes an initial increase of the form $M(t) \sim m_0 \,
t^\theta $, where $\theta$ is a new dynamic exponent. This prediction was
supported by a number of Monte Carlo (MC) investigations \cite{ZHENG}.
These simulations offer also a possibility to determine the conventional
(static and dynamic) critical exponents \cite{LISCZH1,LISCZH2,SCHZHE} 
as well as the critical temperature \cite{SCHZHE}. 
This may eliminate critical slowing down, since the measurements
are performed in the early part of the evolution.

First numerical simulations of the short-time dynamical relaxation
at criticality started from
a high temperature initial state. However, dynamical scaling exists also for
an ordered initial state. No analytical calculations exist for this situation,
but MC investigations were done \cite{ZHENG}. 
All static and dynamic exponents --- expect for 
the new  exponent $\theta$ --- as well as the critical temperature 
can be also obtained with an ordered initial state.

Up to now, simulations of the short-time dynamics
are performed at the critical point 
to determine most of the critical exponents or in vicinity of the 
critical point to determine the exponent $1/\nu z$ and the critical temperature.
However, a further important aspect of simulations is the 
determination of the (spatial) correlation length in 
the long-time regime for temperatures above $T_{\mathrm{c}}$. 
In this letter we investigate the XY model in the high temperature phase,
i.e.\ at finite correlation lengths. Starting from the ordered
state, we measure the autocorrelation function and
from this  we determine the (exponential) autocorrelation time 
$\xi_{\mathrm{t}}$. After that we 
test the predicted scaling behaviour with
the (spatial) correlation length $\xi_{\mathrm{s}}$, where
the values  
for $\xi_{\mathrm{s}}$ are taken
from conventional simulations in equilibrium \cite{WOLFF}.
It is shown, that this relation can be used to calculate
the correlation length (up to a constant), i.e.\
$\xi_{\mathrm{s}}$ is determined 
completely from simulations in the short-time regime. 

The XY model in two dimensions provides a simple model of a system 
with continuous symmetry. It is defined by the  Hamiltonian
\begin{equation}
H= - \beta \,  \sum_{\langle ij \rangle} {\bmf S}_i \cdot {\bmf S}_j \ ,
\end{equation}
where ${\bmf S}_i$ denotes a planar spin of unit 
length at site $i$, $\beta$ is the
inverse temperature ($k$=$J$=1) and the sum
is over all nearest neighbours. This system has an exponential singularity,
i.e.\ the correlation length diverges as 
$\xi_{\mathrm{s}} \sim  \exp ( b \, \tau^{- \nu})$ for 
$\tau=(T-T_{\mathrm{c}})/T_{\mathrm{c}}
\rightarrow 0^+$. This behaviour is different from that of 
a second-order transition, where the correlation length diverges with a power
law. Also, the XY system remains critical below $T_{\mathrm{c}}$,
i.e.\ the spatial correlation function decays algebraically to zero
(quasi-long-range order). This absence of
conventional long-range order for two-dimensional systems
with continuous symmetry was proven by Mermin and Wagner \cite{MERWAG}.
A theoretical description of the transition derived from renormalization group
treatment was given by Kosterlitz and Thouless (KT) \cite{KOSTHO,KOSTER}.
The KT mechanism is based on a topological defect (called vortex) unbinding
scenario.
 
Most of the simulations of the XY model are performed in the thermodynamical
equilibrium. The properties at the critical line 
(e.g.\ critical exponents, critical temperature)
as well as in the high temperature phase (e.g.\ correlation length) 
were examined. In general, simulations at or near the
transition point are affected by the critical slowing down, i.e.\ an increase
of autocorrelation time with 
$\xi_{\mathrm{t}} \sim {\xi_{\mathrm{s}}}^z$.
For some special cases, such as the Ising model or the XY model, this problem
was overcome by the cluster algorithm \cite{CLUSTER1,CLUSTER2}.
Unfortunately, no generalization of
this algorithm  exists  for most of the models (e.g.\ lattice QCD).

An alternative for the determination of critical
exponents is short-time dynamics. 
The power law behaviour in the short-time regime of 
different observables, such
as the magnetization\footnote{Of course, the 
expectation value of the magnetization 
in the XY model in equilibrium at non-zero temperatures is zero.
However, the time dependence  for non-zero initial magnetization
is studied.} or the cumulant
as a function of time, can be used to extract
all static and dynamic critical exponents.
The advantage --- compared to conventional
simulations in equilibrium --- 
is that  critical slowing down is eliminated 
independently from the updating scheme, since
the simulations are performed in the early part of the evolution.
Therefore, in principle this method should work
for any model. For the XY model
the behaviour of the magnetization as a function of time
 starting from an
unordered state was investigated and the critical exponent $\theta$ was
determined \cite{OKSCYAZH}. Also, the time evolution of the magnetization and
the cumulant starting from the ordered state as well as the corresponding
critical exponents were studied \cite{LUOZHE}.
However, no simulations 
in the short-time regime far away from the critical
line were made.

In the following, we investigate the two-dimensional XY model in the high
temperature phase with short-time dynamics. 
We use inverse temperatures of $\beta = 0.82$ - $1.02$,
i.e.\ below the critical point of  $\beta_{\mathrm{c}}\approx 1.1199(1)$ 
\cite{HASPIN}.
We start the relaxation process from the ordered initial state. This
means that the absolute value of the magnetization
\begin{equation}
{\bmf M}(t) = \frac{1}{L^2} \sum_i {\bmf S}_i
\end{equation}
at $t=0$ is one, where $L$ denotes the lattice size.
The system is then released to a dynamic evolution with the Metropolis or
the heatbath algorithm   and the autocorrelation
\begin{equation}
A(t) = \frac{1}{L^2} \left \langle \sum_i {\bmf S}_i(t) \cdot {\bmf S}_i(0)
\right \rangle 
\end{equation}
is measured.
The autocorrelation (of the spins) $A(t)$ is identical to the
autocorrelation of the magnetization 
$\langle {\bmf M}(t) \cdot {\bmf M}(0) \rangle$, 
since all spins in the initial state
are in the same direction. 
Depending on the updating scheme
and $\beta$, the autocorrelation is measured up to $t=40\,000$ MC sweeps.
The lattice size range from $L=128$ for $\beta=0.82$ to $1024$ for
$\beta=1.02$.
In the first case we performed $4000$ independent measurements
(i.e.\ we used different random numbers), while in the
latter case the average was taken over $22$ independent samples.

\begin{figure}
\begin{center}
\mbox{\epsfxsize=6.5cm
\epsfbox{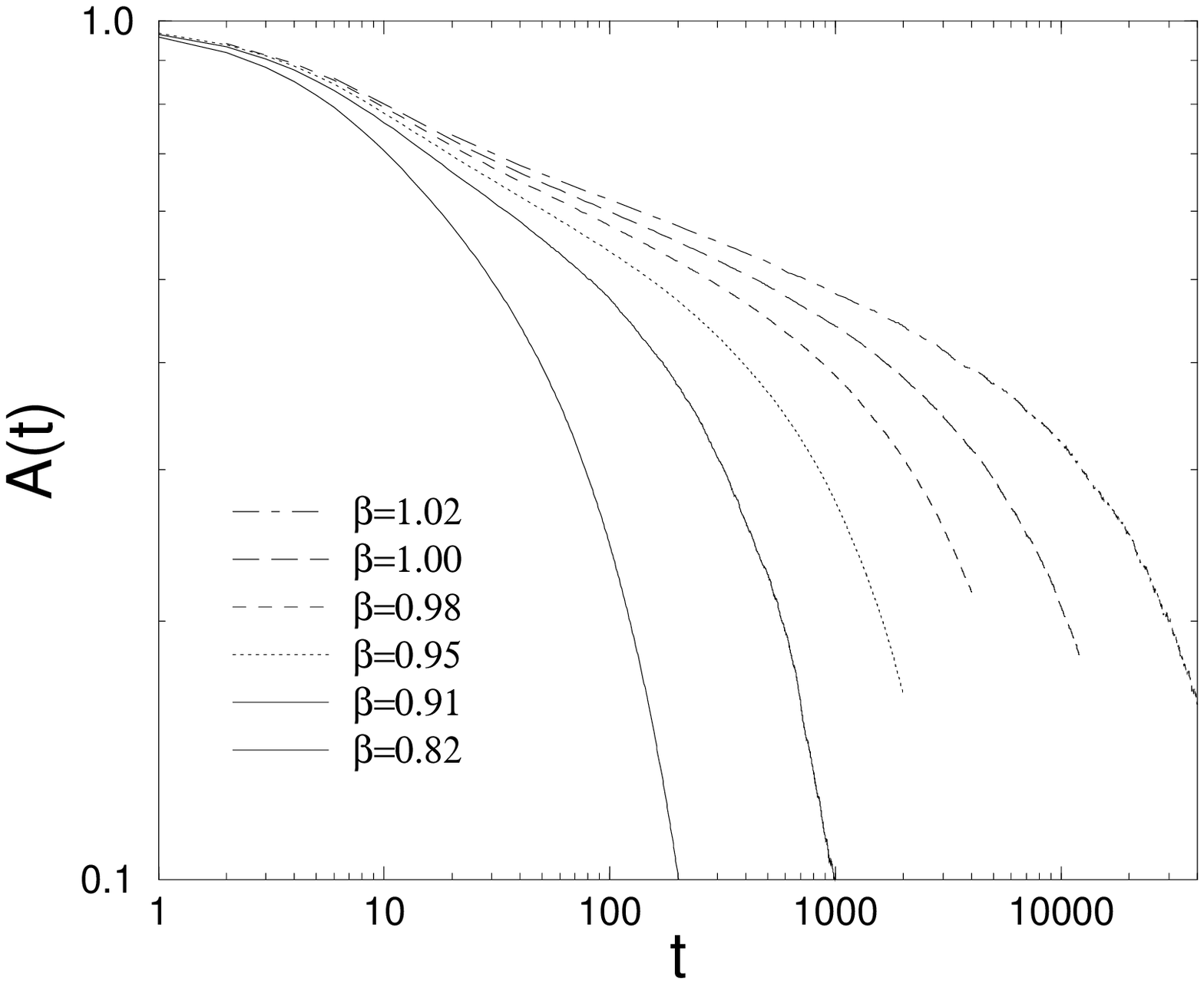}
\epsfxsize=6.5cm
\epsfbox{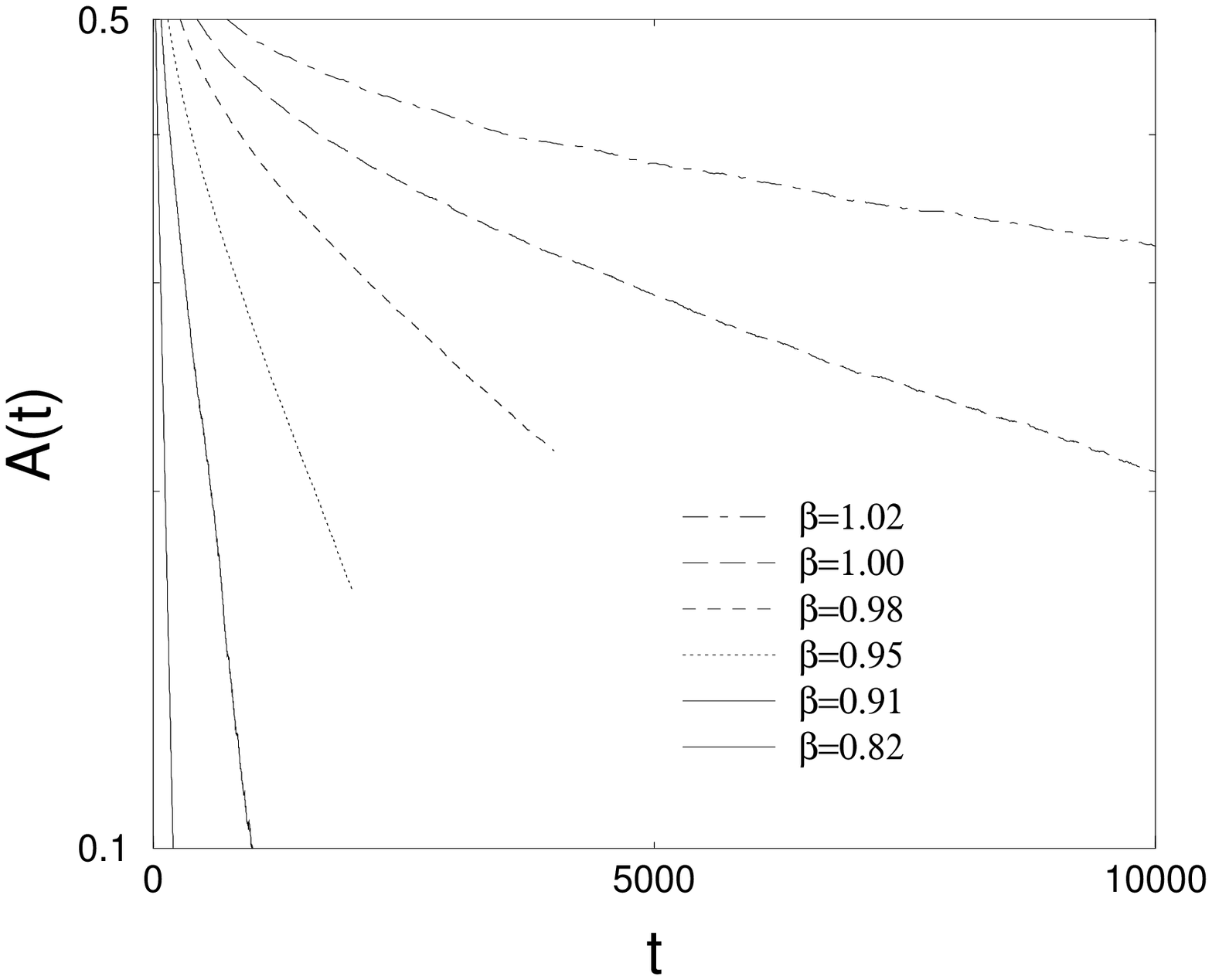}}
\end{center}
{\caption{\label{fig_auto1}
Autocorrelation for different temperatures starting from the ordered state
in (a) a double log and (b) a 
lin-log scale. The MC dynamics is given by the
Metropolis algorithm.
}}
\end{figure}
At the critical line the autocorrelation obeys a
power law \cite{ZHENG}
\begin{equation}
\label{eqacrit}
A(t) \sim t^{-\eta/2 z} \ ,
\end{equation}
while in the high temperature phase the behaviour can be described by
the ansatz
\begin{equation}
\label{eqansatz}
A(t) \sim t^{-\eta/2 z} \, \exp ( -t/\xi_{\mathrm{t}} ) \ .
\end{equation}
The autocorrelation measured for different values of $\beta$ is plotted in
Fig.\ \ref{fig_auto1}, where we used the Metropolis algorithm.
Figure \ref{fig_auto1}(a) shows the autocorrelation in a log-log scale to 
clarify the power law behaviour for small $t$, while the lin-log scale of 
Fig.\ \ref{fig_auto1}(b) exhibits
the exponential decay at larger times. 
Statistical errors of $A(t)$ are of the order of $1 \, \%$.
For microscopic times $t_{\mathrm{mic}}$ up 
to approximately 50 sweeps\footnote{The
exact value depends on $\beta$.
The criterion for the determination of $t_{\mathrm{mic}}$ was that
increasing the value does not lead to an essential
change in the fit parameter or $\chi^2$.
At $\beta=0.81$ we left out $10-80$ sweeps, while we skipped the 
values of the first $20-800$ sweeps at $\beta=1.02$.}
there are deviations from
the ansatz (\ref{eqansatz}). This is similar to simulations at the critical
point \cite{ZHENG}. Therefore, we left out the data for such small $t$,
when we  determined the (exponential)
autocorrelation time $\xi_{\mathrm{t}}$ by fitting the data 
according to Eq.\ (\ref{eqansatz}). In all fits the value of $\eta/2z$
is an independent fit parameter, i.e.\ the value depends on the
temperature. The value decreases from about $0.14$ at $\beta=0.82$
to approximately $0.11$ at $\beta=1.02$.
The autocorrelation is always well described by the ansatz 
(\ref{eqansatz}), i.e.\ the 
$\chi^2$ per degree of freedom is smaller than one. Statistical errors
of $\xi_{\mathrm{t}}$ are obtained by dividing the data into subsamples and
performing the analysis on these independent sets.
The usage of different time intervals 
(i.e.\ different $t_{\mathrm{mic}}$) yields
an estimate for the systematic error.
The quoted error is the sum of
statistical and systematic error. 

\begin{figure}
\begin{center}
\mbox{\epsfxsize=7.5cm
\epsfbox{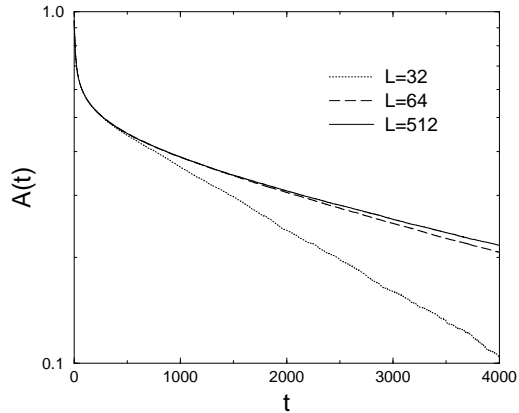}}
\end{center}
{\caption{\label{fig_FSeffects}
Autocorrelation as a function of time at $\beta=0.98$ in a lin-log 
scale for different system sizes. 
}}
\end{figure}

Finite-size effects were studied for
lattice sizes of $L=32$, $64$, $128$, $256$ and $512$ at $\beta=0.98$.
Figure \ref{fig_FSeffects} shows the result for $L=32$, $64$ and $512$.
Only the two smallest lattices show finite-size effects, while
the results for $L=128$ and $256$ coincide
within statistical errors with the data of the largest lattice 
in the time interval $t=[0,4000]$.
For that reason, we omitted these curves in the figure. 
The MC time when a system starts to show finite-size effects
scales with $L^z$. Thus the lattice size of $L=512$ ---
which was used to determine $\xi_{\mathrm{t}}$ 
at $\beta = 0.98$ --- is large enough to 
avoid additional errors coming from a too small system.
Correspondingly, also the lattice sizes for the simulations
at the other temperatures are chosen large enough.

\begin{figure}
\begin{center}
\mbox{\epsfxsize=7.5cm
\epsfbox{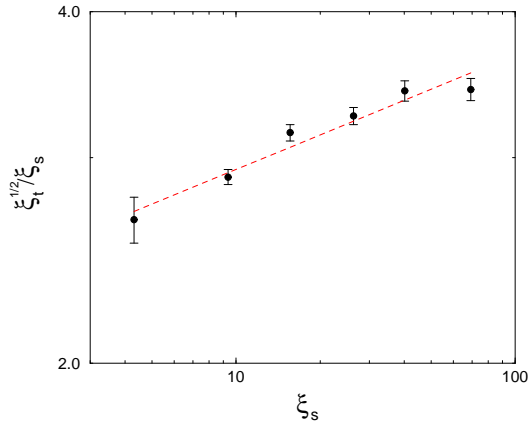}}
\end{center}
{\caption{\label{figxivsxiM}
Test of the scaling relation $\xi_{\mathrm{t}}=c\,{\xi_{\mathrm{s}}}^z$
using the Metropolis algorithm. The slope in the double logarithmic scale
gives the value $z/2-1$.
The dashed line shows a linear fit.
}}
\end{figure}

The autocorrelation
time  $\xi_{\mathrm{t}}$ is related to the spatial correlation length 
(in equilibrium) by
\begin{equation}
\label{eqxiscale}
\xi_{\mathrm{t}}=c \, {\xi_{\mathrm{s}}}^z \ .
\end{equation}
This relation is often used to determine the dynamic critical
exponent $z$ from simulations in equilibrium \cite{XYCSD}
by measuring the  autocorrelation time\footnote{Usually,
in simulations in equilibrium the integrated autocorrelation time
is used.}
as a function of the correlation length.  In the following we 
examine the behaviour of ${\xi_{\mathrm{t}}}$ 
as a function of ${\xi_{\mathrm{s}}}$, where we
have taken the values of $\xi_{\mathrm{s}}$ from Ref.\ \cite{WOLFF}.
In principle we could simply plot ${\xi_{\mathrm{t}}}$ versus
${\xi_{\mathrm{s}}}$ in a double logarithmic scale. However,
in case of a dynamic critical exponent of $z \approx 2$
the behaviour can be seen clearer, if we plot ${\xi_{\mathrm{t}}}^{1/2}/
\xi_{\mathrm{s}}$   for $\xi_{\mathrm{s}} \rightarrow \infty$.
This is done in Fig.\ \ref{figxivsxiM}    
for the MC dynamics with the Metropolis
algorithm in a double logarithmic scale. For a constant
value of $c$ we should see a straight line. The slope gives the
value $z/2-1$. It is visualized by a dashed line and was obtained 
from a linear fit.
The data seem to indicate
that $c$ is indeed a constant and the value of $z$
is $2.19(3)$. 

\begin{figure}
\begin{center}
\mbox{\epsfxsize=7.5cm
\epsfbox{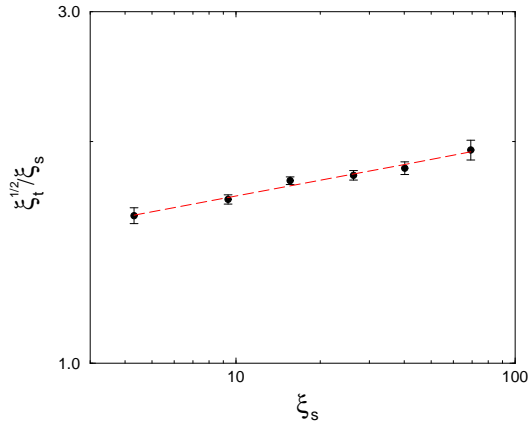}}
\end{center}
{\caption{\label{figxivsxiHB}
Test of the scaling relation $\xi_{\mathrm{t}}=c\,{\xi_{\mathrm{s}}}^z$
for the heatbath algorithm.
}}
\end{figure}

To clarify the situation we also made simulations
using the heatbath algorithm. We expect that corrections 
to the scaling behaviour (\ref{eqxiscale}) far away from the
critical point are smaller than in case
of the Metropolis algorithm. The reason is that
at very high temperatures ($T \rightarrow \infty$,
$\xi_{\mathrm{s}} \rightarrow 0$) the autocorrelation time
$\xi_{\mathrm{t}}$ goes to zero for the heatbath algorithm, while it
approaches a non-zero constant in case of the Metropolis algorithm.
Our results are visualized in Fig.\ \ref{figxivsxiHB}.
Again, we see an almost straight line with non-zero slope of about
$0.07(1)$, i.e.\ $z=2.14(2)$.  Taking both values together we get
$z=2.16(2)$. This value is similar to those of two-dimensional
systems with a second order transition such as the Ising,
3-state Potts or SU(2) model, and with a KT-like transition as the 
6-state clock model \cite{ZHENG}.
Using Eq.\ (\ref{eqxiscale})
with our value of $z$ and $c$,
we could now determine the correlation 
length from  short-time dynamics
by measuring the autocorrelation time $\xi_{\mathrm{t}}$. 

We have seen that short-time
dynamic simulations can be used to determine $\xi_{\mathrm{t}}$.
But also the dynamic critical exponent 
$z$ can be easily extracted from short-time critical dynamics simulations. 
The costs for these measurements are low, because they are not
affected by critical slowing down.
This can be done, e.g., by measuring the dynamic  relaxation 
of the cumulant
\begin{equation}
\tilde{U}(t)=\frac{M^{(2)}(t)}{\left ( {\bmf M }(t)
\right )^2}-1 \ .
\end{equation}
Here $M^{(2)}(t)$ is the second moment of the magnetization.
At the critical point $\tilde{U}(t)$ obeys a power law 
$\tilde{U}(t) \sim t^{-d/z}$,
where $d$ is the dimension of the system. In case of the XY
model this was already done for the Metropolis and the heatbath
algorithm \cite{LUOZHE}. 
However, the result of $z=1.97(4)$ is different from our value.

Short-time critical dynamics
simulations to determine the critical
exponents  are not effected by 
the critical slowing down.
The situation for the measurement of $\xi_{\mathrm{t}}$ in the high 
temperature phase
is different. The MC time which is necessary to determine 
$\xi_{\mathrm{t}}$ is ${\cal O}(\xi_{\mathrm{t}})$, i.e.\
it scales  with ${\xi_{\mathrm{s}}}^z$.
Therefore, the situation is the same as for the simulations 
in equilibrium with
the same algorithms, i.e.\ critical slowing down is not eliminated and
the the CPU time to determine $\xi_{\mathrm{s}}$ scales in the same way.
In our simulations (Metropolis and heatbath algorithm)  
we have a dynamic critical
exponent of $z \approx 2$, but we could also use different
algorithms  and improve the scaling behaviour.
While the scaling behaviour of the CPU time
for the determination of $\xi_{\mathrm{s}}$
is the same for simulations in the short-time regime and in equilibrium, the
total CPU time can be different. 
Also, in the short-time simulation there is no problem to warm-up
the system. Often a large part of simulations in equilibrium 
is only used  to reach the long-time regime.
Of course, in the XY model this problem does not appear, if one
uses the cluster algorithm. However, for many other model
there exists no generalization of this algorithm.

In summary, we have shown for the XY model that 
short-time dynamic relaxation
can be used to determine the (spatial) correlation length. This can be done
by measuring the autocorrelation time and the dynamic critical exponent
with short-time dynamic relaxation and using Eq.\ (\ref{eqxiscale}).
However, at least one value of the correlation length (preferably
near the critical point) is necessary to fix the constant $c$.
These measurements did not fight critical slowing 
down. Nevertheless, the absolute CPU time to measure the correlation length
with short-time simulations can be smaller than the corresponding
simulations in equilibrium.

\begin{ack}
Critical comments on our draft by Lothar Sch\"{u}lke and Wolfgang Bock are
gratefully acknowledged. This work was supported in part
by the Deutsche Forschungsgemeinschaft under Grant No.\ DFG Schu 95/9-1.
\end{ack}

\end{document}